\documentclass[]{spie}  %>>> use for US letter paper
\usepackage{amsmath,amsfonts,amssymb}
\usepackage{graphicx}
\usepackage[colorlinks=true, allcolors=blue]{hyperref}

\title{First on-sky results of ERIS at VLT}

\author[a]{Kateryna Kravchenko}
\author[a]{Yigit Dallilar}
\author[i]{Olivier Absil}
\author[a]{Alex Agudo Berbel}
\author[h]{Andrea Baruffolo}
\author[f]{Markus J. Bonse}
\author[c]{Alexander Buron}
\author[a]{Yixian Cao}
\author[c]{Angela Cortes}
\author[f]{Felix Dannert}
\author[a]{Richard Davies}
\author[d]{Robert J. De Rosa}
\author[a]{Matthias Deysenroth}
\author[g]{David S. Doelman} 
\author[a]{Frank Eisenhauer}
\author[b]{Simone Esposito}
\author[a]{Helmut Feuchtgruber}
\author[a]{Natascha Förster Schreiber}
\author[e]{Xiaofeng Gao}
\author[a]{Hans Gemperlein}
\author[a]{Reinhard Genzel}
\author[a]{Stefan Gillessen}
\author[g]{Christian Ginski} 
\author[f]{Adrian M. Glauser}
\author[c]{Andreas Glindemann}
\author[b]{Paolo Grani}
\author[c]{Pierre Haguenauer}
\author[a]{Johannes Hartwig}
\author[f]{Jean Hayoz}
\author[c]{Marianne Heida}
\author[g]{Matthew Kenworthy} 
\author[c]{Johann Kolb}
\author[c]{Harald Kuntschner}
\author[a]{Dieter Lutz}
\author[a]{Daizhong Liu}
\author[e]{Mike MacIntosh}
\author[d]{Micha{\"e}l Marsset}
\author[i]{Gilles Orban de Xivry}
\author[a]{Hakan Özdemir}
\author[b]{Alfio Puglisi}
\author[f]{Sascha P. Quanz}
\author[a]{Christian Rau}
\author[b]{Armando Riccardi}
\author[a]{Daniel Schuppe}
\author[g]{Frans Snik} 
\author[a]{Eckhard Sturm}
\author[a]{Linda Tacconi}
\author[e]{William D. Taylor}
\author[a]{Erich Wiezorrek}

\affil[a]{Max Planck Institute for extraterrestrial Physics, Gie{\ss}enbachstra{\ss}e 1, 85748 Garching, Germany}
\affil[b]{ INAF-Osservatorio Astrofisico di Arcetri, Largo E. Fermi 5, 50125 Firenze, Italy}
\affil[c]{European Southern Observatory, Karl-Schwarzschild-Str. 2, 85748 Garching, Germany}
\affil[d]{European Southern Observatory, Alonso de C{\'o}rdova 3107, Vitacura, Casilla, 19001 Santiago de Chile, Chile}
\affil[e]{UK Astronomy Technology Centre, STFC, Blackford Hill, Edinburgh, EH9 3HJ, UK}
\affil[f]{Institute for Particle Physics and Astrophysics, ETH Z{\"u}rich, Wolfgang-Pauli-Stra{\ss}e 27, CH-8093 Z{\"u}rich, Switzerland}
\affil[g]{Leiden Observatory, Leiden University, P.O. Box 9513, 2300 RA Leiden, The Netherlands}
\affil[h]{INAF-Osservatorio Astronomico di Padova, Vicolo dell'Osservatorio, 5, 35141 Padova PD, Italy}
\affil[i]{STAR Institute, Universit{\'e} de Li{\`e}ge, All{\'e}e du Six Ao{\^u}t 19c, 4000 Li{\`e}ge, Belgium}

\authorinfo{{kkravchenko@mpe.mpg.de}}

\begin{document} 
\maketitle

\begin{abstract}

ERIS (Enhanced Resolution Imager and Spectrograph) is a new adaptive optics instrument installed at the Cassegrain focus of the VLT-UT4 telescope at the Paranal Observatory in Chile. ERIS consists of two near-infrared instruments: SPIFFIER, an integral field unit (IFU) spectrograph covering J to K bands, and NIX, an imager covering J to M bands. ERIS has an adaptive optics system able to work with both LGS and NGS. The Assembly Integration Verification (AIV) phase of ERIS at the Paranal Observatory was carried out starting in December 2021, followed by several commissioning runs in 2022. This contribution will describe the first preliminary results of the on-sky performance of ERIS during its commissioning and the future perspectives based on the preliminary scientific results.

\end{abstract}


\section{INTRODUCTION}
\label{sec:intro}  % \label{} allows reference to this section

\begin{figure}% [ht]
   \begin{center}
   $\vcenter{\hbox{\includegraphics[height=7cm]{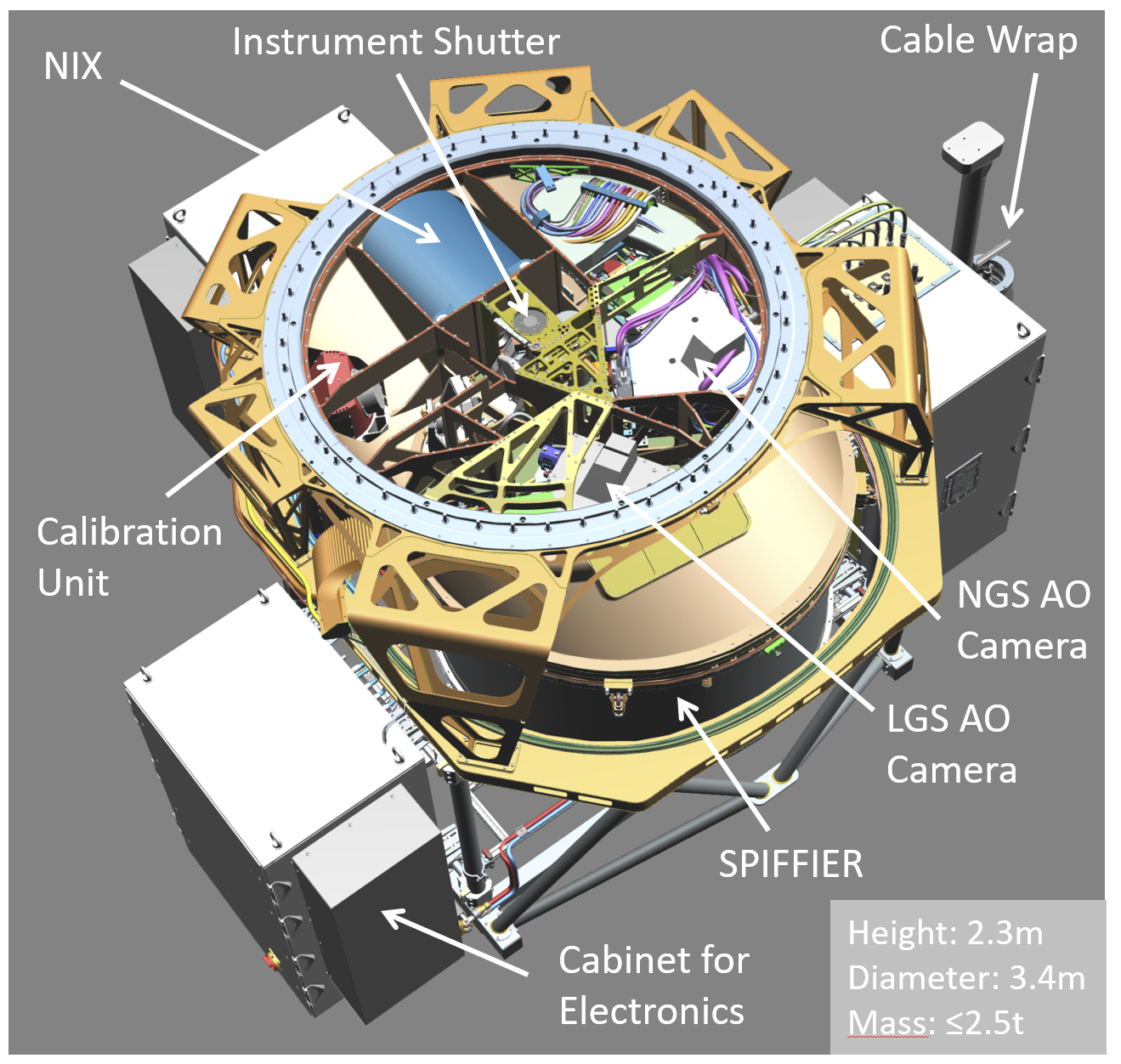}}}$
   $\vcenter{\hbox{\includegraphics[height=6.3cm]{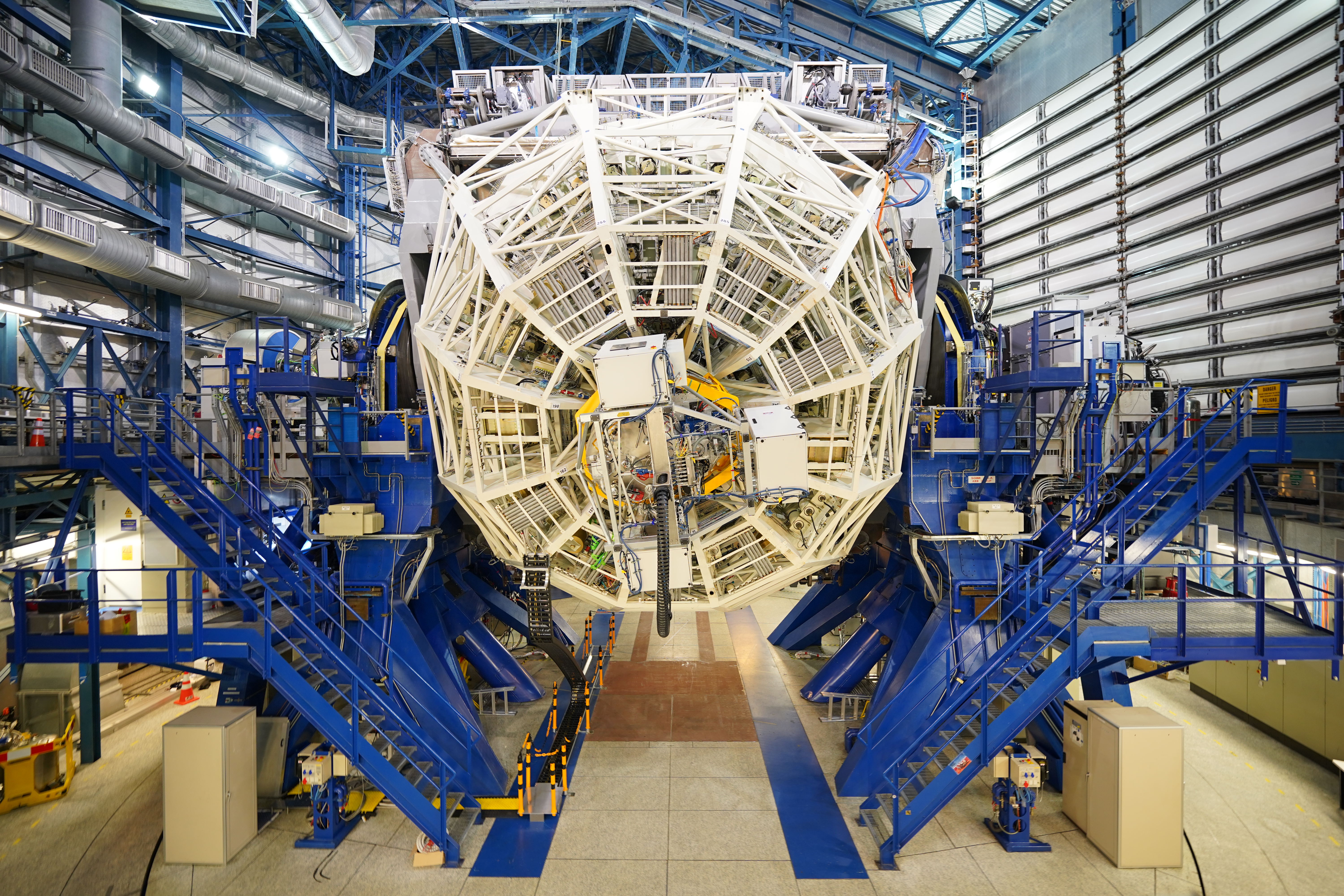}}}$
 \end{center}
   \caption{\textit{Left panel:} The overview of ERIS and its main subsystems: the SPIFFIER spectrograph, the NIX imager, the calibration unit, and the central structure with the LGS and NGS WFS. \textit{Right panel:} ERIS mounted to the Cassegrain focus of VLT-UT4.}
  \label{fig:ERIS_overview} 
   \end{figure}

ERIS is a Cassegrain instrument at the VLT-UT4 of the Paranal Observatory in Chile that will operate at 1-5~$\mu$m \cite{2018SPIE10702E..09D}. It will take over the fundamental adaptive optics (AO) capabilities at the VLT previously provided by NACO and SINFONI and, thus, ensure that the VLT remains at the forefront of AO imaging and spectroscopy into the next decade. The main scientific drivers of ERIS include resolved studies of high-redshift galaxies, astrometry in the Galactic Centre, and characterisation of exoplanets. The ERIS project is being led by a Consortium of Max-Planck Institute for Extraterrestrial Physics (MPE, leading institute), Istituto Nazionale di Astrofisica (INAF Arcetri, Abruzzo and Padova), UK Astronomy Technology Centre (UK-ATC), Institute for Particle Physics and Astrophysics (ETH-Zurich), Netherlands Research School for Astronomy (NOVA Leiden), and European Southern Observatory (ESO). Fig.~\ref{fig:ERIS_overview} displays the overview of ERIS and its main subsystems.

ERIS has two science cameras called SPIFFIER and NIX. SPIFFIER \cite{2016SPIE.9908E..0GG} is an integral field unit (IFU) spectrograph covering the JHK bands, and is an upgraded version of SPIFFI (SPectrometer for Infrared Faint Field Imaging), which was part of SINFONI \cite{2003SPIE.4841.1548E,2004Msngr.117...17B}. SPIFFIER provides simultaneous spectroscopy of 32x64 spatial pixels with a spectral resolution of either $\sim$5000 or $\sim$10000 at three image scales: 25, 100, and 250 mas/px, leading to fields of view (FoV) on the sky of 0.8"x0.8", 3.2"x3.2" and 8"x8". NIX \cite{2016SPIE.9908E..3FP} is an imager operating in the JHK and LM bands and providing a wide range of modes: standard diffraction-limited imaging in JHK (13 and 27 mas/px image scales leading to 26"x26" and 55"x55" FoV, respectively) and LM (13 mas/px image scale leading to 26"x26" FoV) bands, long slit spectroscopy from 3 to 4 $\mu$m and high contrast imaging (HCI) modes from focal/pupil plane coronagraphy to sparse aperture masking interferometry. During science operations, users will select either ERIS/NIX or ERIS/SPIFFIER for their observations. 

The AO module of ERIS provides corrected wavefronts in the J-M bands
to NIX and SPIFFIER and has the following adaptive modes: 

\begin{itemize}
    \item Natural Guide Star (NGS) with an on- or off-axis reference star;
    \item Laser Guide Star (LGS) with an on-axis LGS and off-axis NGS for tip tilt sensing and truth sensing;
    \item Seeing enhancer mode where only the on-axis LGS wavefront sensor (WFS) is used for the high-order correction (in cases when no tip-tilt star is available).
\end{itemize}
\
Since ERIS is mounted on UT4 it makes use of the Adaptive Optics Facility (AOF \cite{2018SPIE10703E..1GO}): one of the four lasers of 4LGSF \cite{hackenberg2016eso} is used to generate an artificial sodium LGS, and the wavefront correction is done by the deformable secondary mirror (DSM) using the Real Time Computer (RTC) platform called SPARTA \cite{2006SPIE.6272E..10F}.   

The calibration data for the SPIFFIER and NIX science observations are provided by the Calibration Unit (CU \cite{2018SPIE10702E..3GD}), which consists of various internal sources for JHK bands: a Quartz-Tungsten Halogen (QTH) lamp for flatfielding, four pencil-ray lamps (Ne, Xe, Kr, Ar) for SPIFFIER wavelength calibration and a Laser Driven Light Source (LDLS) for focusing purposes, position adjustments on the detector and distortion correction (only SPIFFIER). Long-wavelength (LM band) calibrations with NIX are performed exclusively on-sky.  

The AIV phase of ERIS at the Paranal Observatory was carried out between December~2021 and February~2022 followed by its first light and subsequent commissioning, which is still ongoing. The following sections describe technical overview and preliminary performance results of ERIS/SPIFFIER (Sect.~\ref{Sect:SPIFFIER}) and ERIS/NIX (Sect.~\ref{Sect:NIX}) at selected observing modes. The description of the AO sub-system design and preliminary commissioning results are reported in \citenum{AO}.

\section{SPIFFIER}
\label{Sect:SPIFFIER}

\subsection{Instrument overview}

\begin{figure}%[!ht]
    \centering
    \includegraphics[height=9cm]{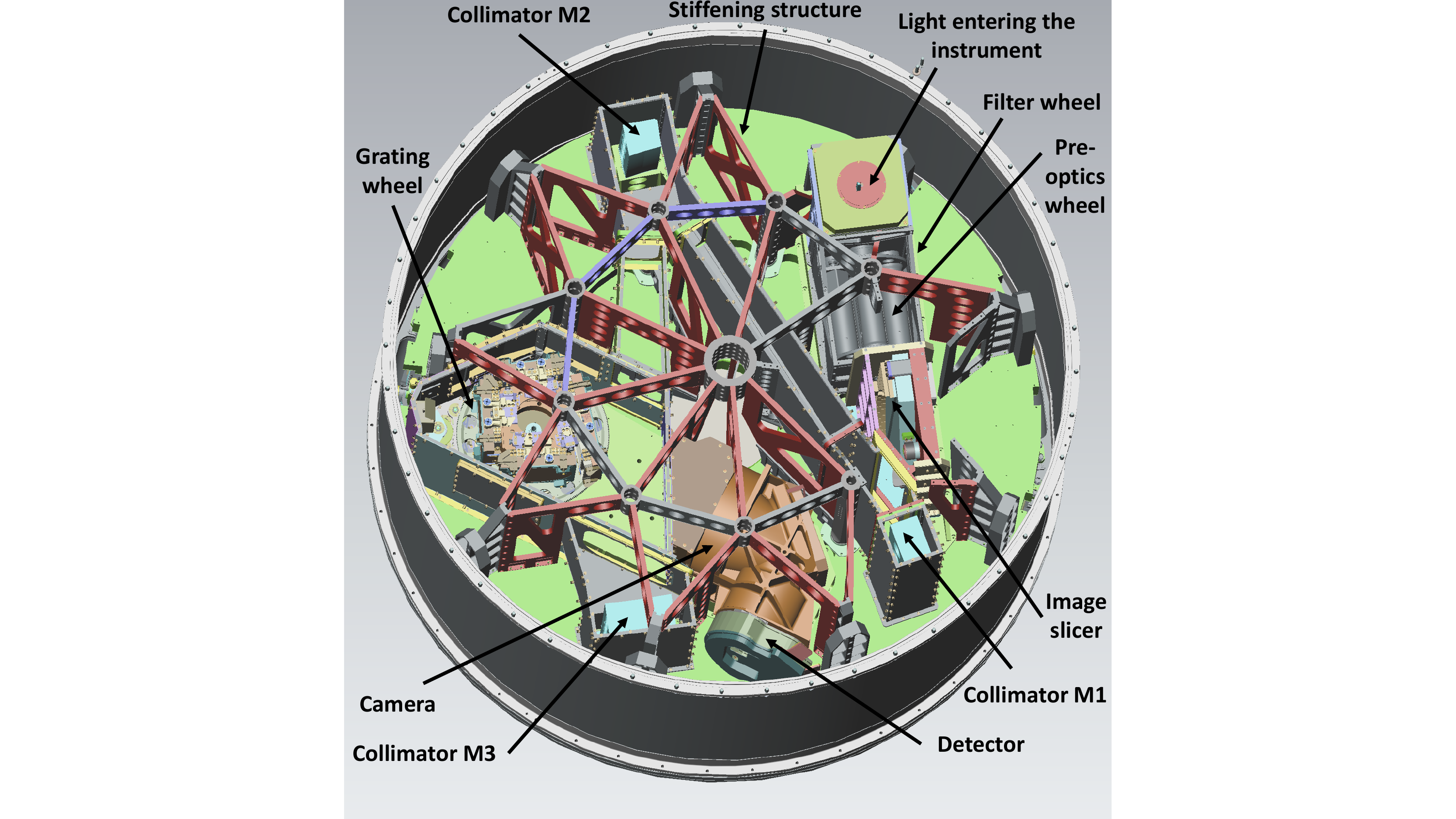}
    \caption{An inside view of the SPIFFIER cryostat with indications for the locations of various opto-mechanical elements. Some housing covers and parts of the stiffening structure are not shown for better illustration.}
    \label{fig:spi_ins}
\end{figure}

SPIFFIER is the integral field spectrometer and is an upgraded/refurbished version of SPIFFI featuring a new HAWAII 2RG (2x2k) detector and four new gratings (J, H, K, high-resolution(J,H,K)) providing better spectral resolution thanks to more symmetric and narrower line spread functions. Addition of the high-resolution grating leads to twice higher spectral resolution compared to the nominal gratings. SPIFFIER provides simultaneous spectra of 32x64 spatial pixels (spaxels) at three image scales: 250, 100, and 25 mas/px, leading to field of views on the sky of 8"x8", 3"x3", and 0.8"x0.8", respectively. Figure~\ref{fig:spi_ins} shows an image of the open SPIFFIER cryostat with indications for the locations of the various elements. All components are cooled in a bath cryostat to the temperature of liquid nitrogen ($\sim$77 K). The liquid nitrogen reservoir sits below the instrument base plate.

\begin{figure}%[!ht]
    \centering
    \includegraphics[width=0.99\textwidth]{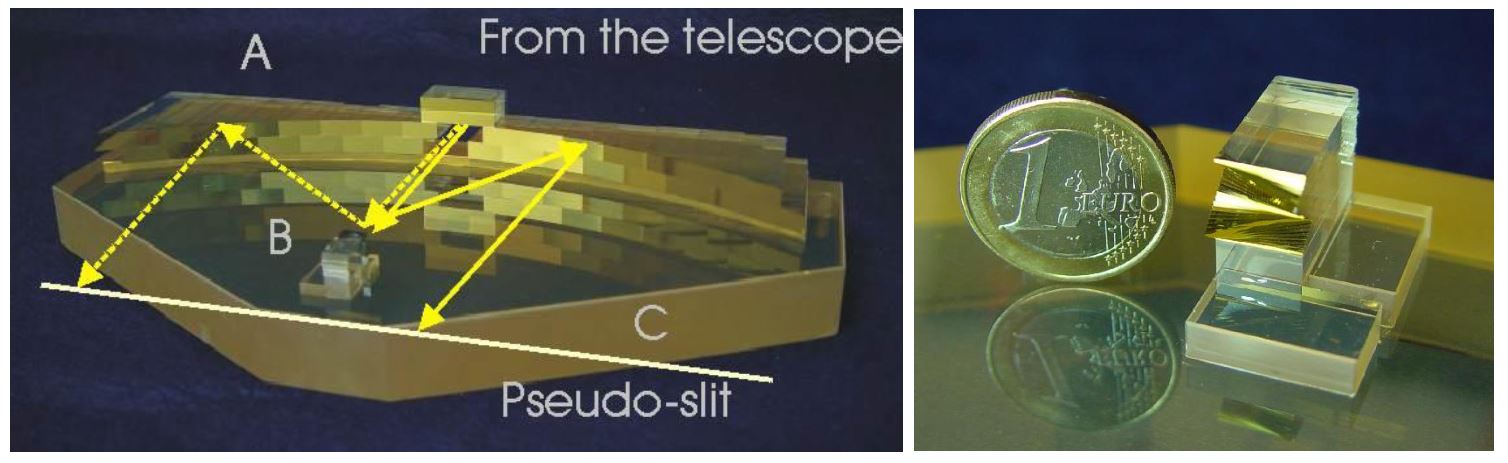}
    \includegraphics{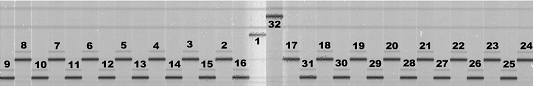}
    \caption{\textit{Top panels:} SPIFFIER image slicer (adapted from \citenum{2004SPIE.5492.1123I}). The small image slicer B (shown on the top right panel) cuts the image into stripes and reflects them onto the big image slicer A to create a pseudo long slit to be fed into spectrometer. Both image slicer components are mounted to a baseplate C. \textit{Bottom panel:} The layout of the slitlets on a raw detector frame.}
    \label{fig:spi_slicer}
\end{figure}

The light enters from the top. Below the entrance focal plane baffle, a triplet lens unit collimates the light onto a cold stop for the suppression of the thermal background. Just in front of the cold stop is the motorized filter wheel housing the band-pass filters. After the cold stop, the motorized optics wheel provides the interchangeable lens systems for the three different image scales: 25, 100 and 250 mas/px. The light of the pre-optics is focused on the image slicer: a stack of 32 small plane mirrors – the so-called small slicer – slices the image and redirects the light towards the 32 mirrors of the big slicer, which rearranges the slitlets to a long pseudo-slit, which appears as a brick-wall pattern on the detector. All parts are of Zerodur and are optically contacted (without using any glue). Each one of the 32 slitlets is imaged onto 64 pixels of the detector. Figure~\ref{fig:spi_slicer} shows the image slicer and the positions of the slitlets on a raw SPIFFIER frame. The slitlets run horizontally across the imaging field-of-view and are numbered from top to bottom on the small slicer. 

\begin{figure}%[!ht]
    \centering
    \includegraphics[width=0.47\textwidth]{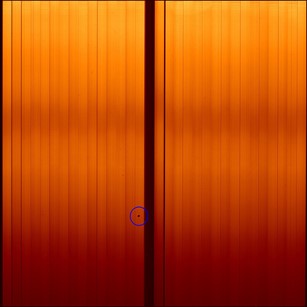}
    \caption{SPIFFIER K-band flat at 25 mas pixel scale. A clump of cold pixels is marked with circle.}
    \label{fig:spi_badp}
\end{figure}

After the image has been sliced and re-arranged into a pseudo-slit, three diamond turned mirrors (M1, M2 and M3 in Fig.~\ref{fig:spi_ins}) collimate the light onto the gratings. The first mirror is spherical, and the other two have an oblate elliptical shape. All mirrors are made from aluminum and are gold-coated for higher reflectivity. In total, four gratings are implemented on the grating drive. They are based on Zerodur blanks ruled into a gold layer on the reflecting surface. Three of the gratings cover the J (1.1-1.4 $\mu$m), H (1.45-1.85 $\mu$m), and K (1.95-2.45 $\mu$m) spectral bands at a resolution of $\sim5000$ superior to the SINFONI gratings by a factor of $\sim$1.3 in K to $\sim$2.5 in J. The fourth grating is the high-resolution grating and replaces the previous R$\sim$1500 H+K grating of SINFONI. This grating doubles the spectral resolution in a given band but reduces the wavelength range by a factor of two. For each band, users can select either the short, middle or long wavelength regime. A five-lens camera system then focuses the spectra on the detector. All lenses have a multi-layer anti-reflection coating optimized for the wavelength range from 1.05-2.45~$\mu$m. 

A detailed description of the SPIFFI instrument design can be found in \citenum{2003SPIE.4841.1548E}, with part of the upgrades to SPIFFIER described in \citenum{2017JATIS...3c5002G}.

\subsection{Detector} 

The old Hawaii 2RG detector of SPIFFI was replaced by a new Hawaii 2RG detector because of better persistence and cosmetics. The new detector was delivered from Teledyne Imaging Sensors. The SPIFFIER detector operates using the up-the-ramp readout scheme with a frame time of 1.6 seconds. The conversion gain is near 2~e-/ADU and the read noise is 12~e- rms at the shortest exposures. The minimum noise of $\sim$7~e- rms is reached around 80~s exposures. The dark current amounts to 0.19~e-/s.

The SPIFFIER detector has randomly distributed bad pixels. These can be interpolated over during data reduction. The only defect worth noting is a clump of cold pixels (not sensitive to light) about 10 pixels in diameter, marked with a dark blue circle in Fig.~\ref{fig:spi_badp}. This spot falls into slitlet 16 in all configurations, i.e. in the middle of the reconstructed image.

   \begin{figure}% [ht]
   \begin{center}
   \includegraphics[height=7cm]{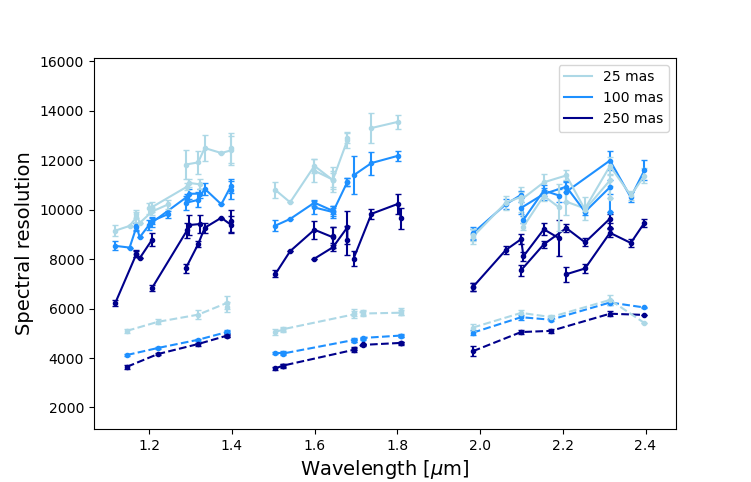}
   \end{center}
   \caption{Spectral resolution as a function of wavelength for the low- (dashed lines) and high-resolution (solid lines) JHK gratings of SPIFFIER. Colors correspond to the three pixel scales (25, 100, and 250 mas). Errorbars represent standard deviations from the averages over 32 slitlets.}
  \label{fig:resolution} 
   \end{figure}

\subsection{Performance}

\subsubsection{Spectral resolution}

For a given SPIFFIER grating, the spectral resolution can be calculated using wavelength calibration data provided by the Ne, Xe, Kr and Ar penray lamps of ERIS CU. The procedure is to fit a Gaussian to individual spectral lines in a single lamp exposure and divide the wavelength of a spectral line by the FWHM of its Gaussian ﬁt. Using pipeline-processed wavelength calibration maps, the wavelengths and widths of various spectral lines were extracted for all gratings and pixel scales. The calculated spectral resolution values are illustrated in Fig.~\ref{fig:resolution}. SPIFFIER provides spectral resolution of about 5000 and 10000 for the low- and high-resolution JHK grating configurations, respectively. The resolution increases for smaller pixel scales and longer wavelengths.

   \begin{figure}% [ht]
   \begin{center}
   \includegraphics[height=6cm]{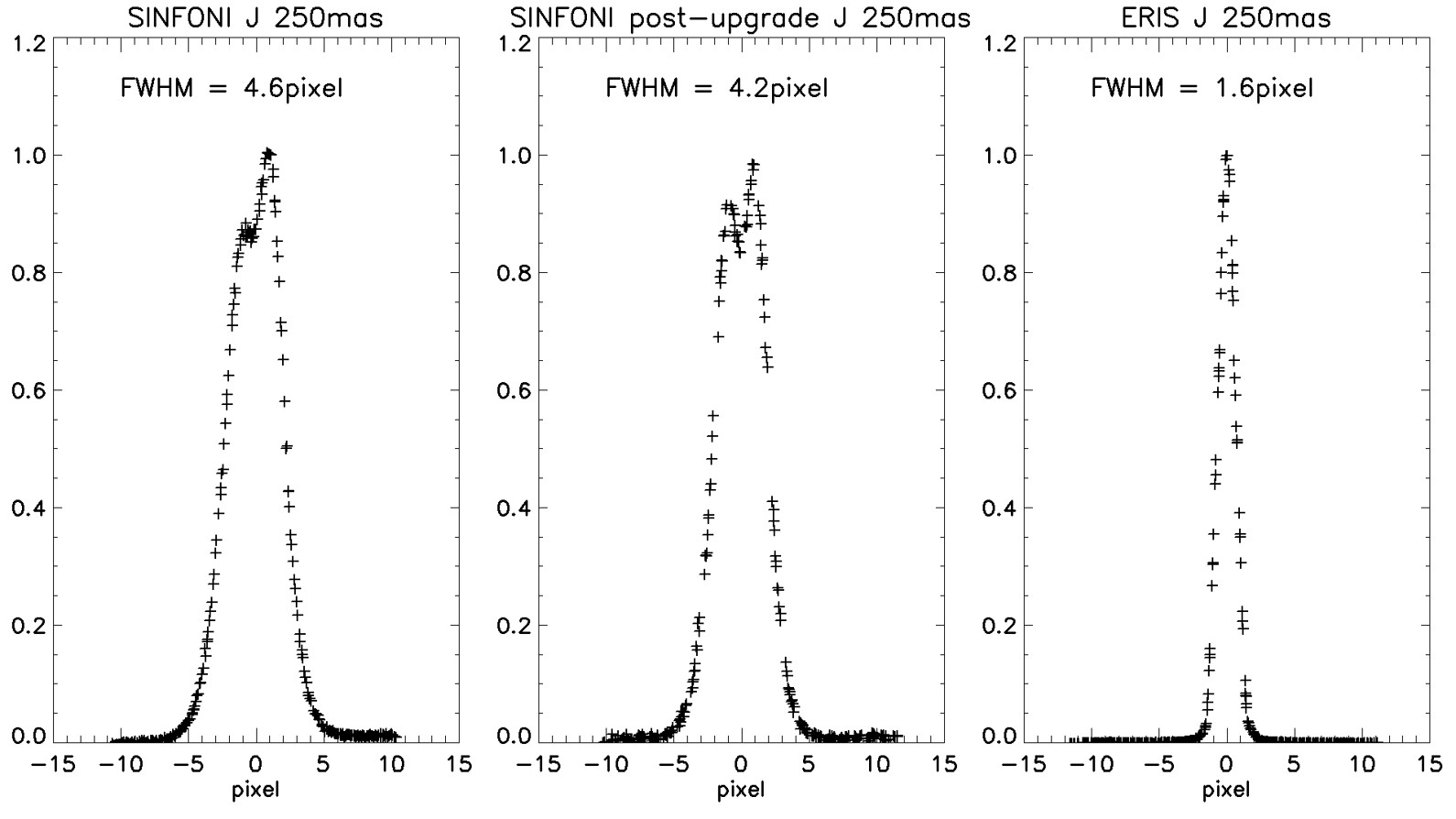}
   \end{center}
   \caption{Improvement of instrumental line profile shapes from asymmetric seen in SPIFFI (left, center) to symmetric with the new gratings of SPIFFIER (right) for the same  spectral resolution configuration.}
  \label{fig:line_profiles} 
   \end{figure} 

\subsubsection{Instrumental line profiles}

The instrumental line profiles of SPIFFI were characterized by asymmetric shapes deviating from a Gaussian shape (left panel of Fig.~\ref{fig:line_profiles}). The SPIFFI gratings were made of NiP coated aluminum with lightweighting. It was found that the interplay of the lightweighting structure with the stress induced by the NiP coating at cryogenic temperatures caused deformations of the grating surface and led to degraded instrumental line profiles \cite{2017JATIS...3c5002G}. 

The new gratings of SPIFFIER are based on Zerodur blanks without any lightweighting and, therefore, substantially improve the shapes of the spectral lines. Since line profiles are undersampled on the detector (the widths are less than two pixels), an approach similar to \citenum{2012SPIE.8448E..09T} was used to obtain hypersampled spectral line proﬁles for the detailed line-shape analysis. A series of penray lamp calibration exposures with CU was taken for each band and pixel scale. For a series of exposures of a particular band and pixel scale, the grating wheel was turned by a few encoder positions between each exposure, which corresponds to a shift of the central wavelength on the detector by approximately 0.1 pixels. These slightly shifted exposures are referred to as "babysteps". In total, 21 babysteps exposures per band and pixel scale were obtained. The combination of the babysteps exposures allows to create hypersampled line profiles for a given instrument configuration. Fig.~\ref{fig:line_profiles} illustrates an example of the resulting oversampled SPIFFIER line profile compared to those from SPIFFI. The instrumental line profiles of SPIFFIER are symmetric in all bands thanks to the improved design of its diffraction gratings.   

\section{NIX}
\label{Sect:NIX}

\subsection{Instrument overview}

\begin{figure}%[ht]
   \begin{center}
   \includegraphics[height=6cm]{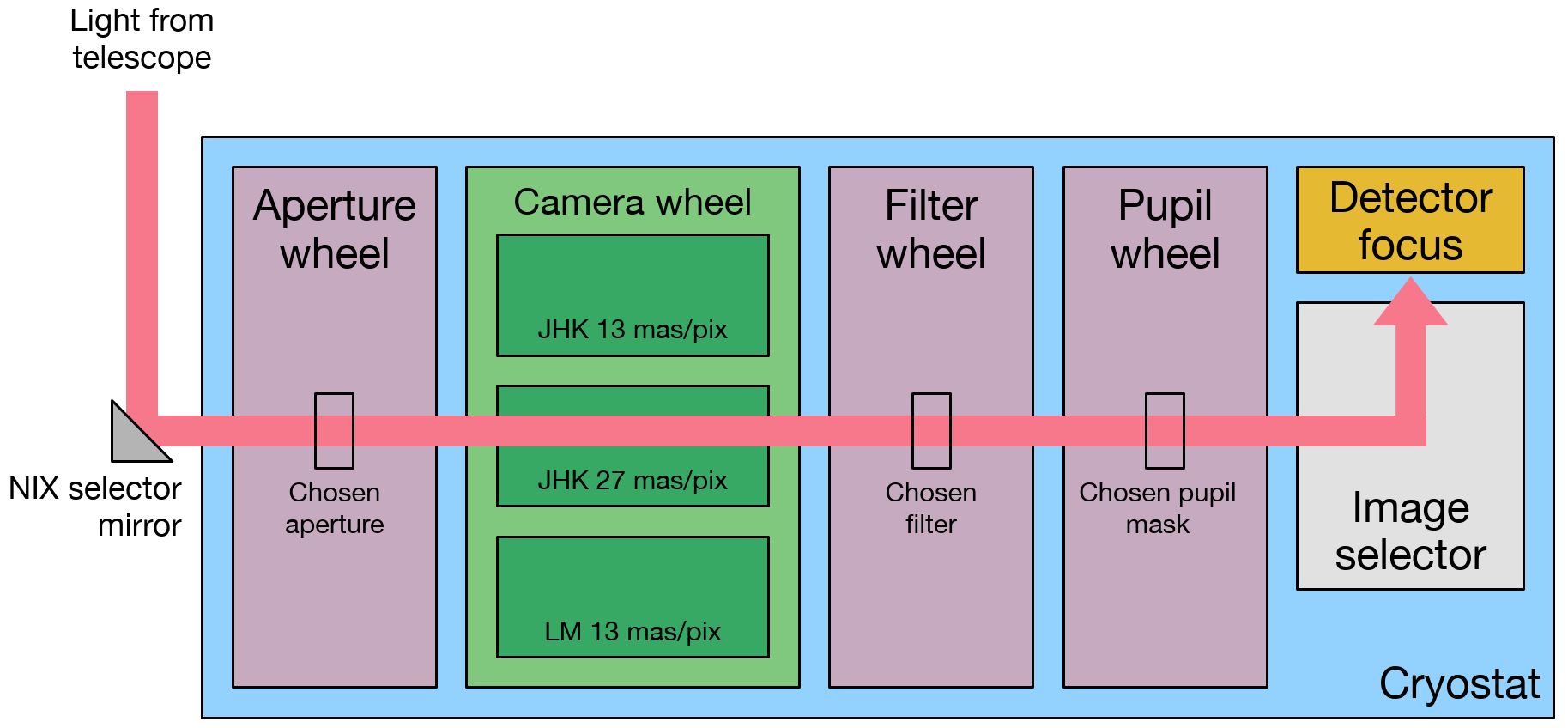}
   \end{center}
   \caption{A sketch of the light path through the NIX cryostat.} 
  \label{fig:nix_layout} 
   \end{figure}

\begin{figure}%[ht]
   \begin{center}
   \includegraphics[height=9cm]{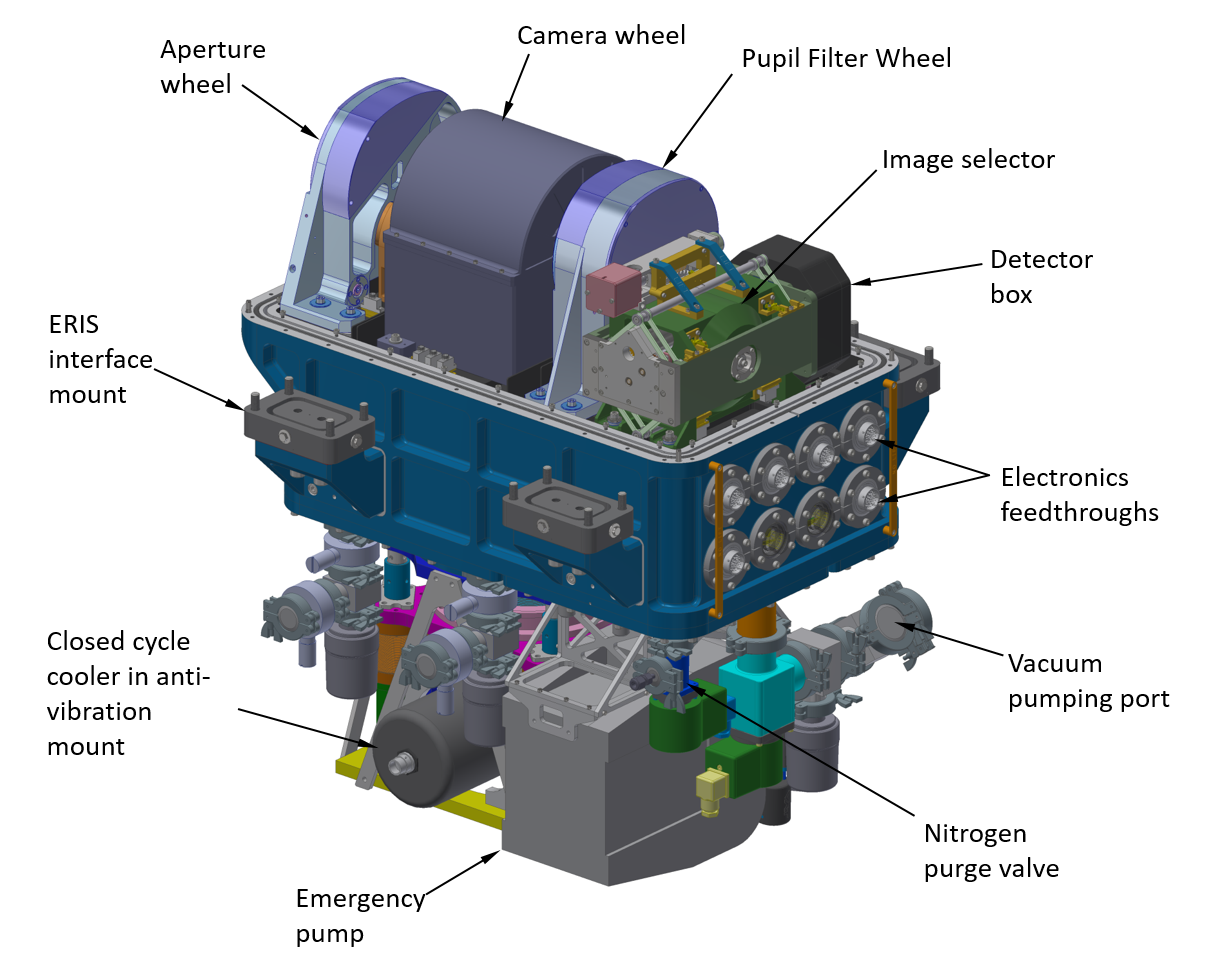}
   \end{center}
   \caption{A three-dimensional view of NIX instrument.} 
  \label{fig:nix_layout_3d} 
   \end{figure}

The NIX imager provides diffraction-limited imaging capabilities in J-M bands (from 1 to 5~$\mu$m); focal plane coronagraphy with Annular Groove Phase Mask (AGPM) in L-M bands\cite{Mavet2013}; pupil plane coronagraphy with grating vector Apodizing Phase Plate (gvAPP) in K-M bands\cite{2021ApOpt..60D..52D}; sparse aperture masking (SAM) in J-M bands; and long-slit spectroscopy (LSS) in L-band (from 3 to 4~$\mu$m). The primary elements of NIX are indicated in Figs.~\ref{fig:nix_layout} and \ref{fig:nix_layout_3d}. Light enters NIX via the NIX selector mirror. This selector mirror is part of the ERIS system and, when deployed, directs the light from UT4 into the NIX imager instead of the SPIFFIER spectrograph. Inside NIX, the light passes through the aperture window (Calcium Fluoride) into the NIX cryostat, indicated by the blue region in Fig.~\ref{fig:nix_layout} (and also marked in Fig.~\ref{fig:nix_layout_3d}). Nearly all components and mechanisms inside the cryostat are cooled to 75K to limit thermal background radiation; the detector is the only item cooled further to $\sim$35K by a closed cycle cooler.

After the cryostat window the light passes through the aperture wheel located at the telescope focal plane. It houses various field masks (including a blank position) that are used depending on the observing mode; these are interchangeable by means of a deployment mechanism driven by a stepper motor providing high positional repeatability. The design and performance of the other wheels is very similar. The next mechanism is the camera wheel, which contains three different camera barrels. The camera designs are optimized to use the minimum number of elements to maximize the throughput, while being as axially compact as possible. The camera lenses are fabricated from Barium Fluoride, IRG2 and Zinc Selenide. Two cameras are optimised for the shorter wavelengths (J, H and K), providing spatial scales of 13 mas/px or 27 mas/px. The third camera barrel is for the longer wavelengths (L and M) delivering 13 mas/px. 

From the camera wheel the light passes through the filter and pupil wheels, which are identical mechanisms housed within a single unit. Both wheels can house up to 18 elements that can be combined in various ways for the different operating modes of NIX. The filter wheel houses all the optical filters. The pupil wheel contains a variety of elements that include pupil masks, a grism and additional filters. A set of fold mirrors (the image selector) then brings the light to a focus on the detector. The image selector has four positions: one for each of the three cameras and one to allow pupil imaging. The light is then detected by a Teledyne Hawaii-2RG 5~$\mu$m cutoff detector which is read out using the standard ESO NGC controller. The detector focus stage is used to adjust internal focus of the NIX detector. 

\subsubsection{Detector}

The NIX detector allows two readout configurations for the user; a slow up-the-ramp configuration with a frame time of 1.6 seconds and a fast uncorrelated configuration with fastest full frame readout speed of 30 Hz. The former is configured with an effective gain of 5.2 e-/ADU. The read noise is 19 e- rms at the shortest exposures, dipping around 9 -e rms near 100 second exposures and the dark current dominates from here onwards with 0.1 e-/s. The latter is configured with an effective gain of 2.6 e-/ADU and the read noise per read is about 50 e-/s. 

The detector configuration comes with several caveats such as odd-even channel effect, for which a fix using the reference pixels is included with the pipeline. The clumps of bad pixels have the most notable impact in the design of observations. The users can determine a specific position angle on-sky to perform their observations to position poorly affected regions to the uninteresting regions and/or a design a dither pattern minimizing the overlap of bad pixels. Note that aside from these regions, the cosmetics are more typical, $<1\%$. For example, high contrast imaging modes are demonstrated to operate well on the upper half of the detector.

\begin{figure} %[ht]
   \begin{center}
   $\vcenter{\hbox{\includegraphics[width=0.45\textwidth]{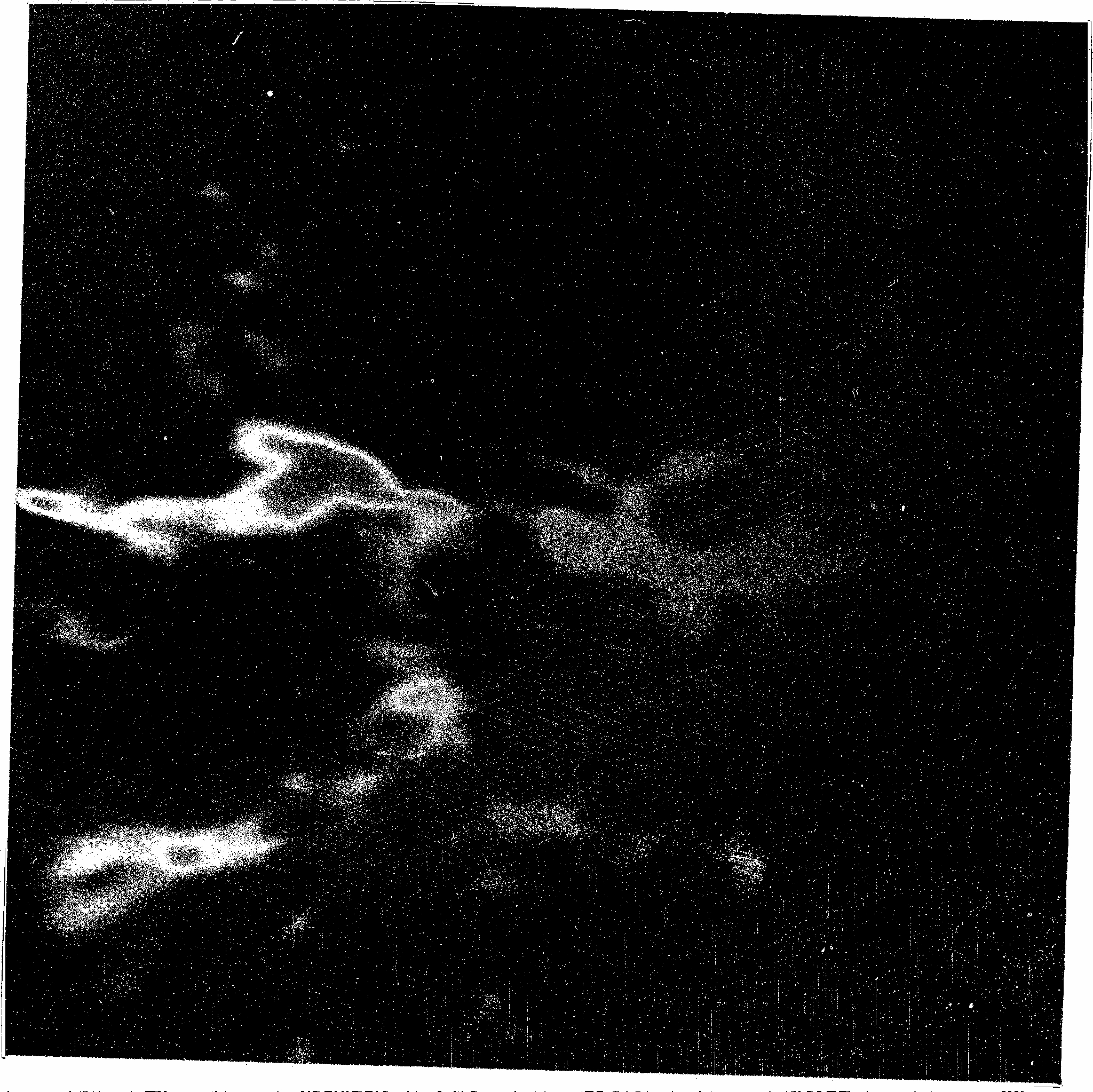}}}$
   $\vcenter{\hbox{\includegraphics[width=0.45\textwidth]{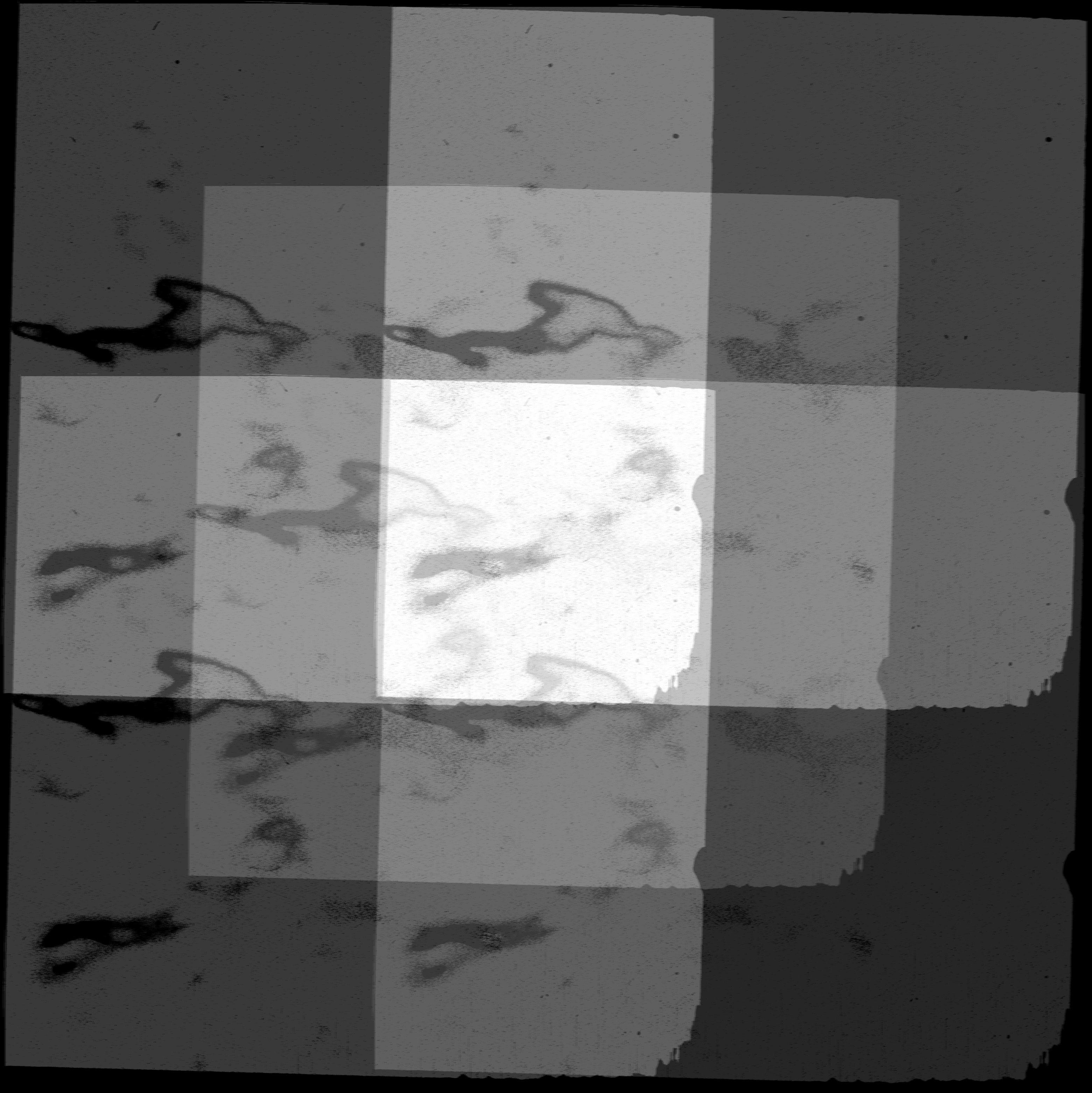}}}$
   \end{center}
   \caption{Left: Bad pixel map derived with the pipeline. Right: Confidence map in the stacked final product for a dither pattern with five offset positions.}
  \label{fig:BP} 
\end{figure}

Such an example is given in Figure \ref{fig:BP} with a choice of five point dither pattern on a 14" box using the 13mas-JHK camera. This certainly reduces the effective exposure time in the regions where there is overlap with the clump of bad pixels. However, the NIX pipeline delivers stacked variance and confidence maps for users to properly assess the performance of each pixel in the final product.

\subsection{Performance}

\subsubsection{Astrometric Calibrations - Omega Centauri}

After a preliminary coordinate calibration with the telescope offsets observing an isolated star, we opted for Omega Centauri crowded field observations for astrometric calibrations of the instrument from there onward. Such an observation provides many benefits in a single attempt taking advantage of GAIA sources sampled in the field. We can accurately determine the position angle, pixel scale of each camera and the precise pointing of each frame. Hence, this analysis also serves as input for the coordinate calibration of stages of the ERIS AO system, similarly a feedback for SPIFFIER as its FOV is small in comparison to perform such accurate analysis.

\begin{figure} %[ht]
   \begin{center}
   $\vcenter{\hbox{\includegraphics[width=0.45\textwidth]{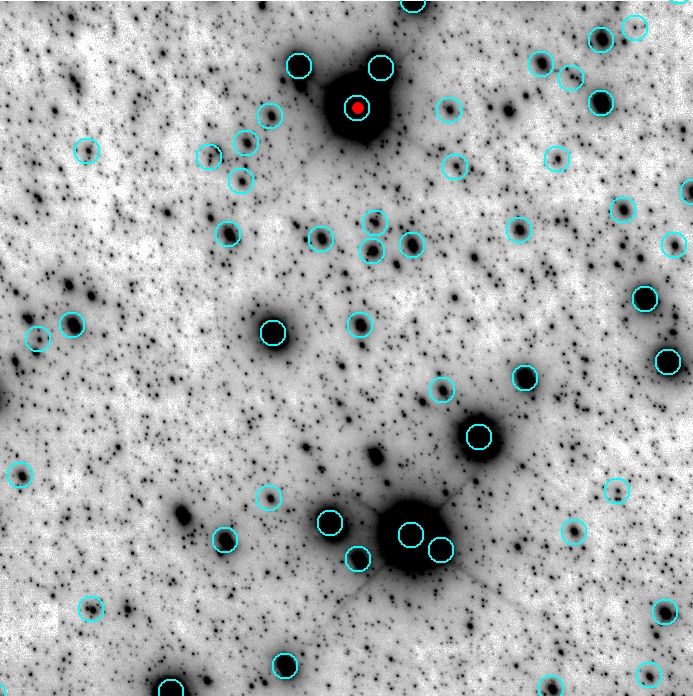}}}$
   $\vcenter{\hbox{\includegraphics[width=0.45\textwidth]{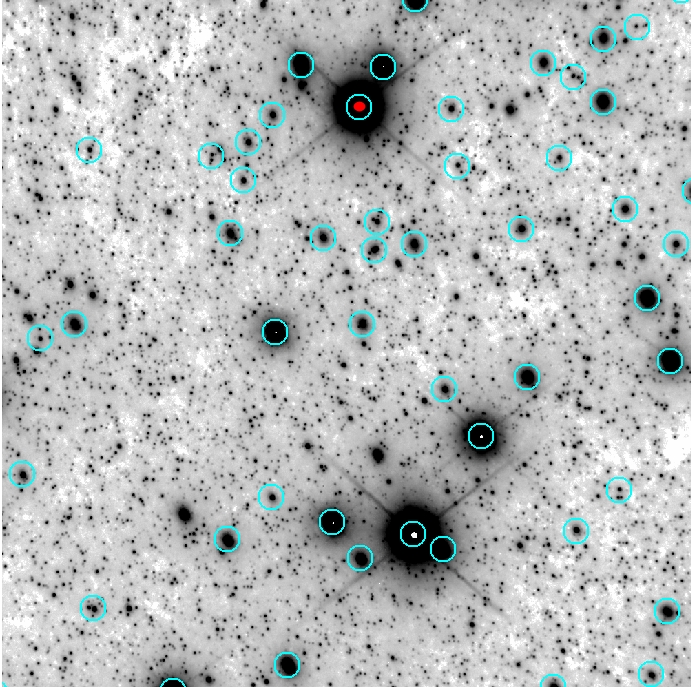}}}$
   \end{center}
   \caption{Combined image around our original pointing near the center of Omega Centauri (27"x27" mas FOV); the left image with the 13mas/px and the right image with the 27mas/px camera. Red dot marks the tip-tilt star (2MASS J13264631-4728402, I=$\sim$11 mag) and available GAIA sources are overlaid with cyan.}% \textcolor{red}{Can do this more cleverly...}}
  \label{fig:OmegaCen} 
\end{figure}

We observed the same field of Omega Centauri as depicted in Figure \ref{fig:OmegaCen} in two separate occasions. Our first visit in April 2022 was with sub-optimal AO performance as we were in the earlier stages of the commissioning. However, the data were useful for the development and testing of our astrometric calibration routines and ERIS observing blocks. In return, this provided an early feedback for the NIX pipeline and ERIS AO system, and served as a reference for our second visit in July 2022. In the July observations, we derived a position angle of -0.4$\pm$0.05 degrees and pixel scale of 13.09$\pm$0.008 mas/px for the 13mas-JHK camera out of 25 frames, and +1.5$\pm$0.02 degrees and 27.92$\pm$0.009 mas/px for the 27mas-JHK camera out of nine frames. The values inside the parentheses indicate the 1-$\sigma$ deviation among the frames.

\begin{figure} %[ht]
   \begin{center}
   \includegraphics[width=0.45\textwidth]{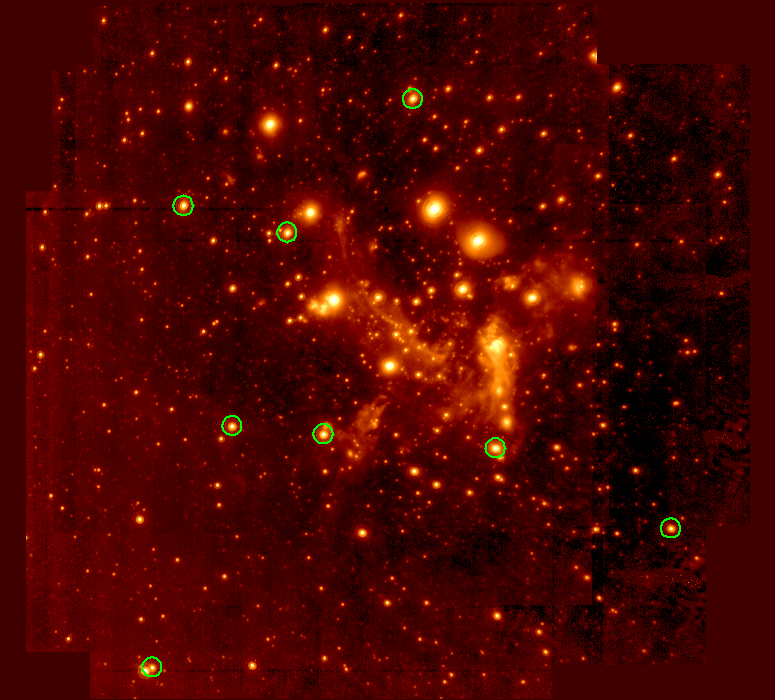}
   \end{center}
   \caption{Stacked auto-jitter images of NIX Galactic Center observations in L$'$ band. The markers denote the SiO masers, the trajectory of which are traced to current date and used for astrometric calibration.}
  \label{fig:GC_Lp} 
\end{figure} 

The fields are also selected on purpose in order to match the pointing of recent HST observations. Hence, this will also serve as accurate distortion characterization for both cameras, for which the analysis is still in progress.

\subsubsection{Astrometric Calibration - Galactic Center}

For the astrometric calibration of the 13mas-LM camera, we observed the Galactic Center relying on the accurately known trajectories of SiO masers\cite{2007ApJ...659..378R}. Ten diffraction-limited images with sufficient quality were used for this purpose. The stacked image is given in Figure \ref{fig:GC_Lp} with the eight SiO masers used in astrometric calibration depicted with circles. Our analysis resulted in position angle of -0.4 degrees and pixel scale of 13.03 mas/px. Poor weather conditions hampered further observations of the Galactic Center and of Omega~Cen as additional reference field.

%Due to the combination of poor weather conditions and Omega Centauri being a lesser known territory in comparison to the Galactic Center in longer wavelengths, we considered this as a safer choice with a probability of visiting Omega Centauri later during the July run. However, the weather conditions prevented the latter. Even the Galactic Center observations were rather eventful due to weather conditions. We were able to acquire about 10 diffraction limited images out of 25. 

\subsubsection{High-contrast imaging with gvAPP}

High contrast imaging is used to image faint circumstellar material or planetary companions, where the diffraction halo of the central star is the dominant noise source in the regions of interest. NIX provides focal and pupil plane coronagraphy as well as sparse aperture masking modes. This section is focused solely on the pupil plane coronagraphy for which the first preliminary on-sky data were obtained. 

\begin{figure} %[ht]
\begin{center}
   $\vcenter{\hbox{\includegraphics[width=0.45\textwidth]{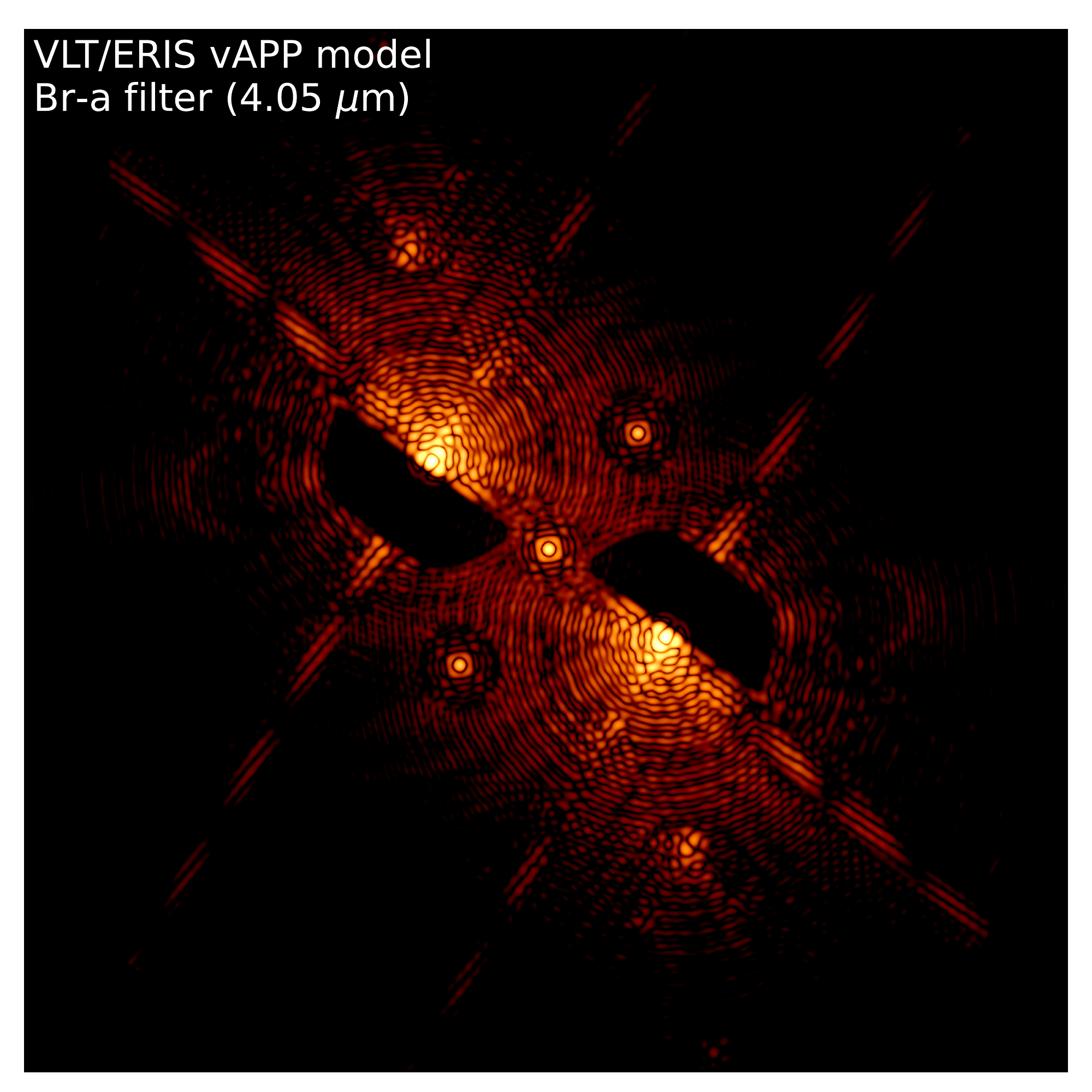}}}$
   $\vcenter{\hbox{\includegraphics[width=0.45\textwidth]{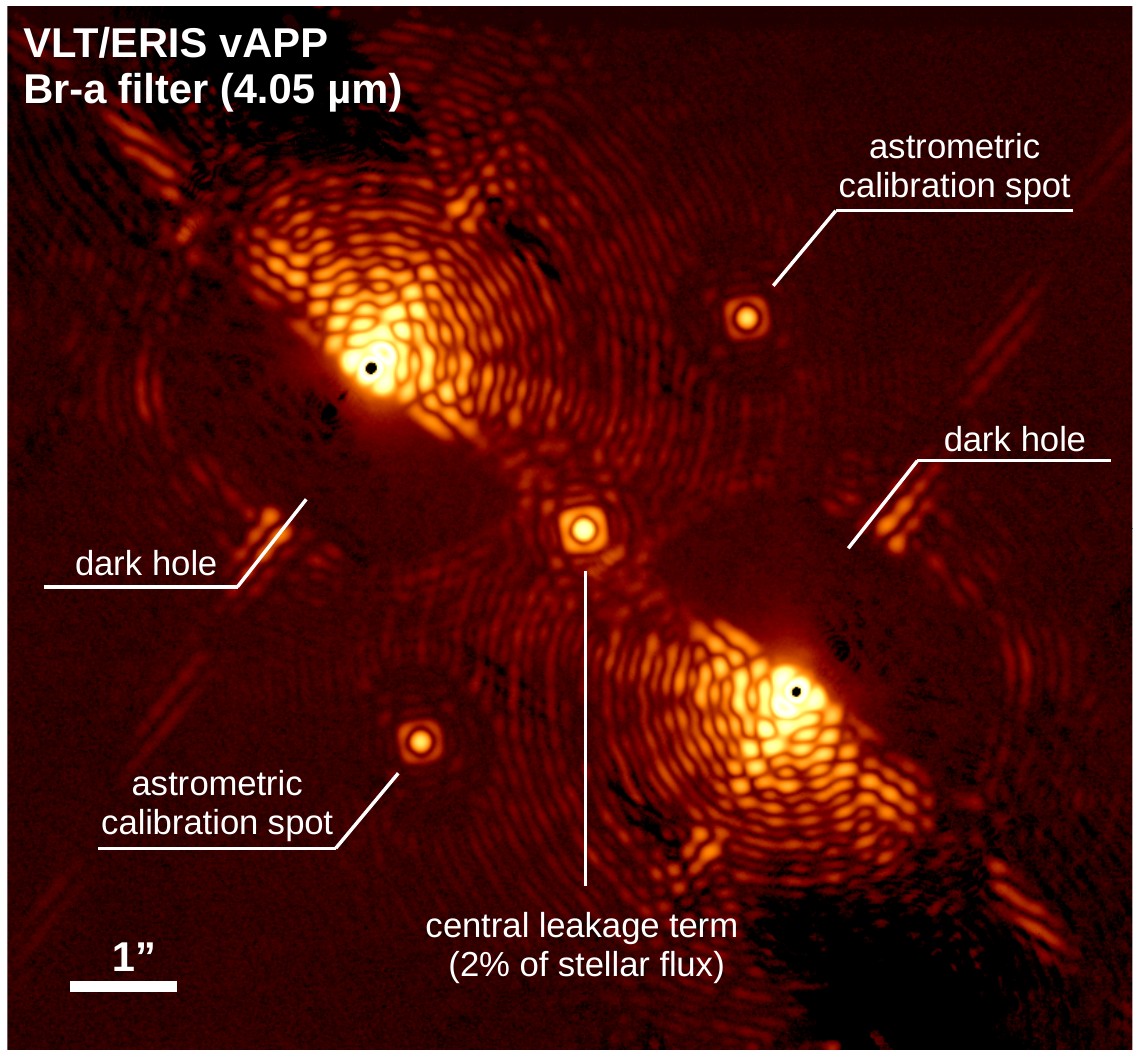}}}$
    \end{center}
   \caption{\textit{Left panel:} Theoretical PSF obtained with the gvAPP mask in Br-$\alpha$ filter of NIX, showing the central faint reference PSF and the two main PSFs with opposite D-shaped high contrast region. \textit{Right panel:} First on-sky PSFs of the standard star $\iota$~Cap with the gvAPP in the Br-$\alpha$ filter.}
  \label{fig:gvAPP} 
\end{figure}

The pupil plane coronagraph is a grating vector Apodizing Phase Plate (gvAPP \cite{2021ApOpt..60D..52D}). The gvAPP is a single optic that goes in the pupil plane of the telescope. It adds a phase pattern across the wavefront of the whole telescope pupil and therefore modifies the PSF of any point source in the field of view as illustrated in Fig.~\ref{fig:gvAPP}. 
%The gvAPP generates three PSFs of a star (see Fig.~\ref{fig:gvAPP}). 
The central PSF (referred to as the "leakage" PSF) is the PSF of the telescope pupil. % and appears in the same position as the directly imaged PSF. 
It contains about 2\% of the transmitted flux and acts as a photometric and astrometric reference. Two additional PSFs appear on either side of the leakage PSF, with the remaining large fraction of the stellar flux split evenly between the two. Each of these two PSFs has a dark D-shaped region of suppression on one side of the star. The two D-shaped regions of both images are on opposite sides of the star, and so provide nearly 360 degrees of suppression with an approximate inner working angle of three $\lambda/D$ when combined.

We observed the bright (2.2~mag in L) standard IR star $\iota$~Cap in July~2022 using gvAPP in the Br-$\alpha$ filter. In data pre-processing, the significant near-infrared  background noise is calibrated by subtracting pairs of nodding positions of the target on the detector.  The right panel of Fig.~\ref{fig:gvAPP} illustrates the observed PSF, which is almost identical to the theoretical PSF calculated for the same instrument configuration (left panel of Fig.~\ref{fig:gvAPP}): both show the central faint reference PSF and the two main PSFs with opposite D-shaped high contrast region. 

\begin{figure} %[ht]
\begin{center}
   \includegraphics[width=0.45\textwidth]{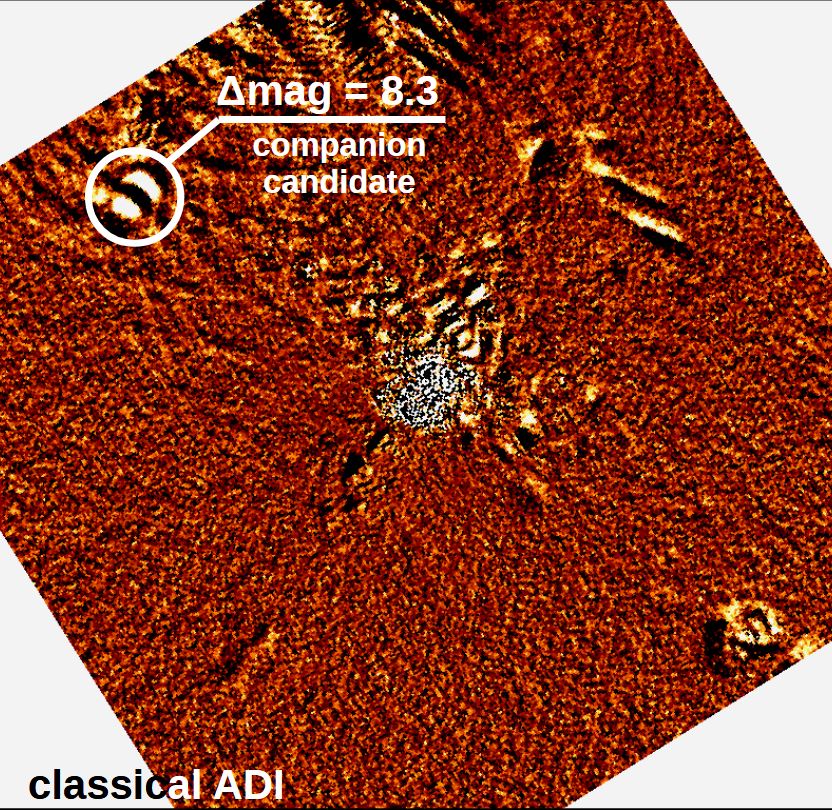}
   \end{center}  
   \caption{The ADI-processed image of $\iota$~Cap in the Br-$\alpha$ filter. White circle marks the location of the detected point source. %\textit{Right panel:} Contrast curve for 900-second exposure with gvAPP in the Br-$\alpha$ filter of NIX. The black dashed line is the contrast curve for one of the two dark D shaped PSFs. The red solid line is the contrast curve after subtracting the opposite gvAPP PSF.}
   }
  \label{fig:companion} 
\end{figure} 

\begin{figure}%[!ht]
    \centering
    \includegraphics[width=0.85\textwidth]{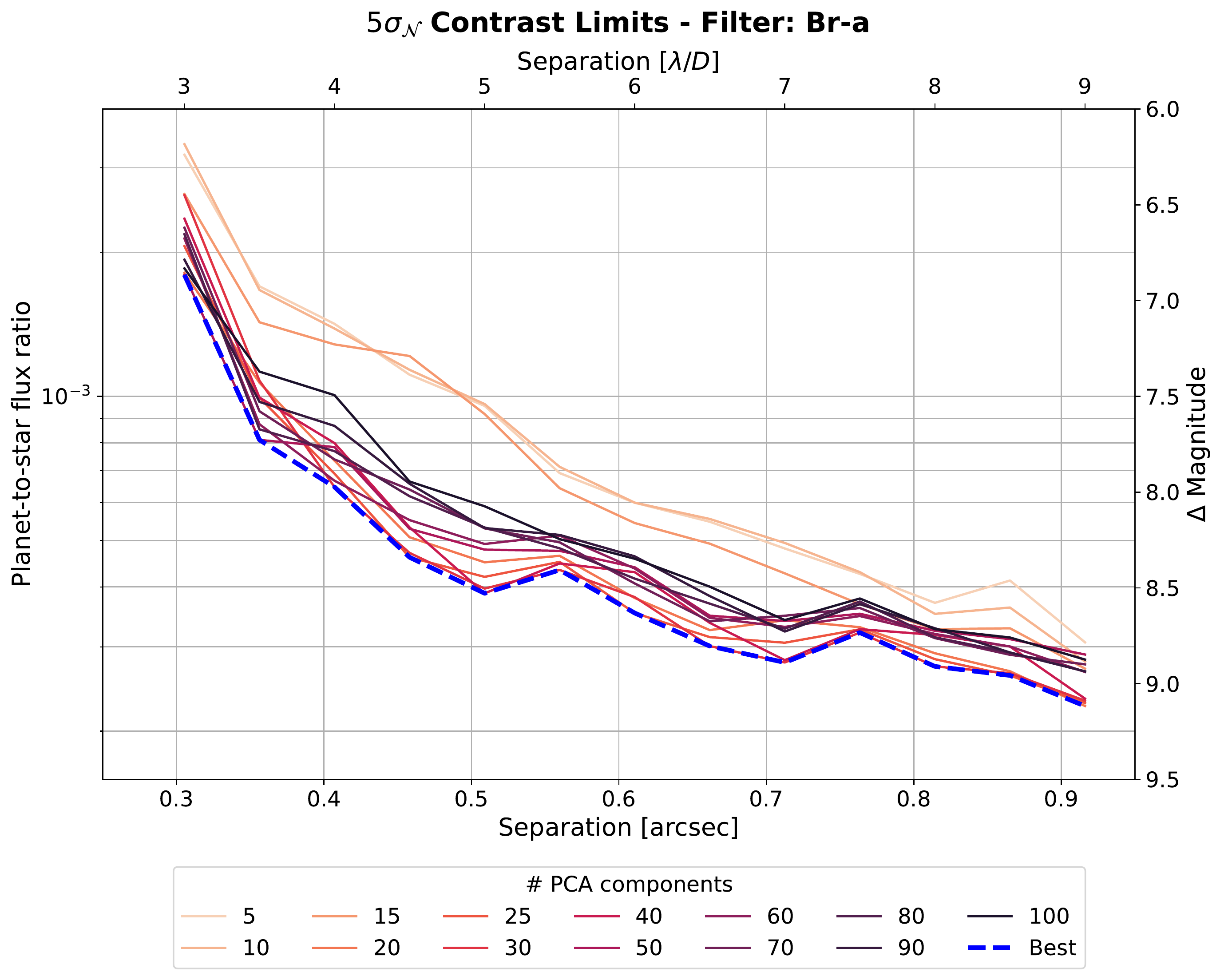}
    \caption{Contrast curve corresponding to a 5$\sigma$ detection for a 735~s exposure with the gvAPP in the Br-$\alpha$ filter of NIX. The blue-dotted curve indicates the deepest achievable contrast under choice of the optimal number of principal components. Observing conditions were good with an average seeing of 0.6~arcsec and average  precipitable water vapor (PWV) of 1.2~mm.}
    \label{fig:nix_app_cnt_bra}
\end{figure}

\begin{figure}%[!ht]
    \centering
    \includegraphics[width=0.85\textwidth]{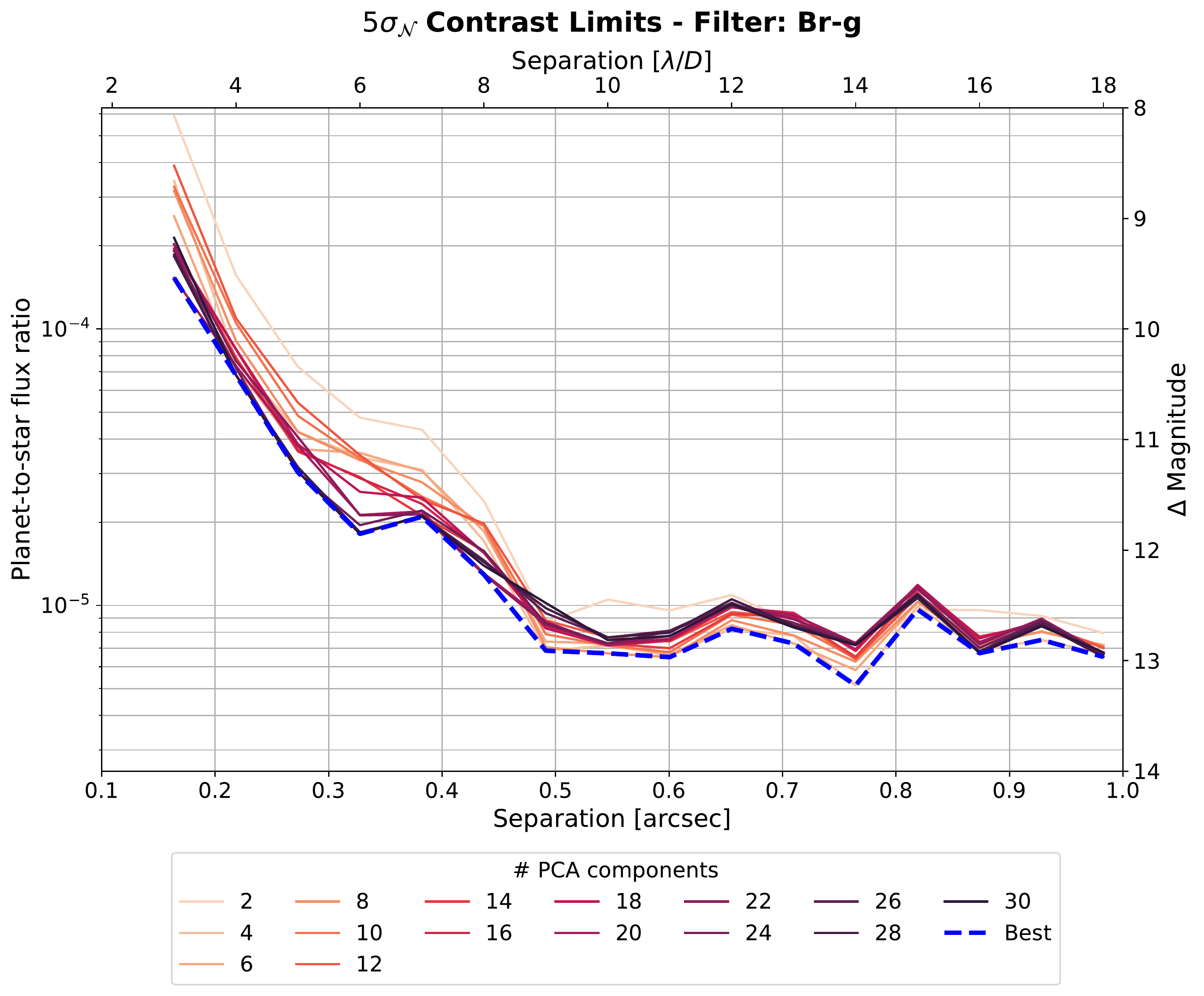}
    \caption{Contrast curve corresponding to a 5$\sigma$ detection for a 2700~s exposure with the gvAPP in the Br-$\gamma$ filter of NIX. The blue-dotted curve indicates the deepest achievable contrast under choice of the optimal number of principal components. Observing conditions were fair with an average seeing of 0.9~arcsec and average PWV of 6.9~mm.}
    \label{fig:nix_app_cnt_brg}
\end{figure}

As next step, the data were post-processed by applying classical angular differential imaging (ADI), which takes advantage of the rotation of the sky field during the observation (since the telescope tracks the pupil rather than the field) to remove the PSF of the star (which is static on the detector), while a companion astrophysical source (e.g. exoplanet) moves on the detector from one exposure to the next. Fig.~\ref{fig:companion} illustrates the resulting PSF-subtracted image of $\iota$~Cap with the first detection of a point source with separation of 1.6" and a contrast of roughly 8.3 mag relative to $\iota$~Cap in the Br-$\alpha$ filter. With the age of $\iota$~Cap (about 0.5 Gyr) the detected point source is either a low mass M-dwarf companion or a background star. 

Fig.~\ref{fig:nix_app_cnt_bra} illustrates an L-band contrast curve of the gvAPP in the Br-$\alpha$ filter using the $\iota$~Cap dataset. The results are based on fake planet injections at different positions and brightnesses around the host star. We use the unsaturated regions of the APP PSF as the fake planet signal. The data were reduced using ADI-based Principle Component Analysis (PCA) \cite{10.1111/j.1365-2966.2012.21918.x} and detection limits are determined by using the package described in \citenum{2022BAAS...54e.392B} (assuming Gaussian noise). The resulting curve identifies which planets can confidently be rejected after 735~s of integration time. By conducting a longer observation (1 hour), one can expect the contrast limits to deepen by about one magnitude.

To demonstrate the contrast performance at shorter wavelengths, we observed the K=3.45~mag star $\gamma$~Gru with the gvAPP in the Br-$\gamma$ filter in July 2022 for a total integration time of 45~min. The data were processed in the same manner as described for the Br-$\alpha$ filter, and the resulting contrast curve is shown in Fig.~\ref{fig:nix_app_cnt_brg}. The contrast curves for other NIX filters will be measured during forthcoming commissioning runs. 

%Fig.~\ref{fig:companion} illustrates an L-band contrast curve of the gvAPP in the Br-$\alpha$ filter using the $\iota$~Cap dataset. The black dashed line is the contrast curve for one of the two dark D shaped PSFs, and the red solid line is the contrast curve where the opposite gvAPP PSF is used to subtract residual speckles generated within the instrument optics. 900~s of the on-sky integration leads to a contrast of 11~mag and is expected to go down by about 1~mag with longer (e.g. 1 hour) exposures. The contrast curves for the remaining NIX filters will be measured during forthcoming commissioning runs.  

\section{Science outlook}

\subsection{Galactic Center}

SINFONI/SPIFFI was built speciﬁcally for observations and studies of the Galactic Center. This allowed major scientific breakthroughs in the field, for example, the discovery of young B stars (the so-called "S-stars") in close orbits around the SgrA$^*$ black hole \cite{2005ApJ...628..246E}. By monitoring the orbits of these stars it was possible to derive the mass and the distance to the central black hole \cite{2009ApJ...692.1075G}. SPIFFI also led to the discovery of infrared flares of SgrA$^*$ and characterization of their spectra \cite{2005ApJ...628..246E}. The Galactic center with its supermassive black hole is an excellent target to test general relativity in the strong-ﬁeld limit, which was done by the peri-center passage of the star S2 \cite{2009ApJ...707L.114G}. Finally, long-term monitoring of the Galacic center with SPIFFI led to the discovery of the gas cloud G2 falling towards the black hole \cite{2013ApJ...763...78G}. 

\begin{figure}% [ht]
   \begin{center}
   $\vcenter{\hbox{\includegraphics[height=7cm]{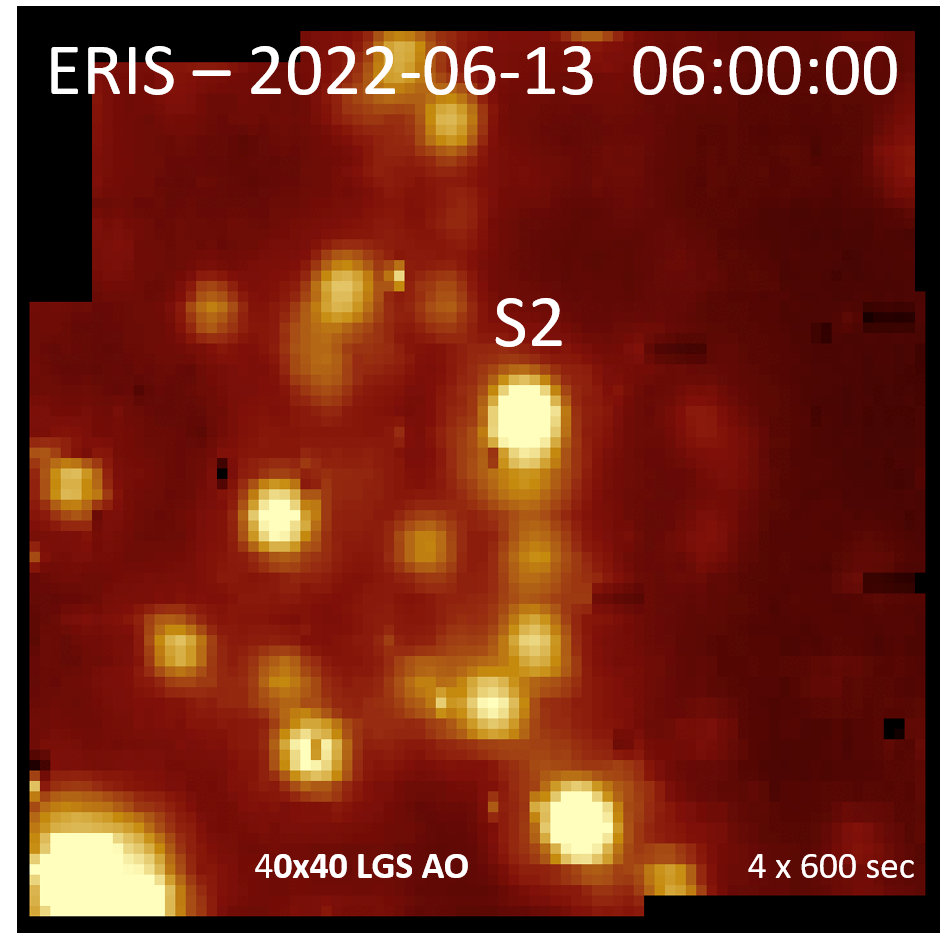}}}$
   $\vcenter{\hbox{\includegraphics[height=7cm]{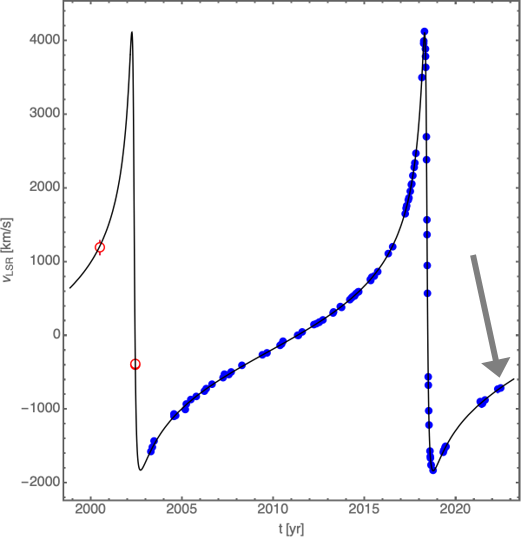}}}$
 \end{center}
   \caption{\textit{Left panel:} Stacked  SPIFFIER reconstructed image of the central arcsecond of the Galactic center in the Br-$\gamma$ line (K~band). The spatial scale is 25mas/px. The location of the star S2 is marked. The integration time was 40~min on source. \textit{Right panel:} Radial velocity (RV) variations of the star S2 during last 20 years observed with different instruments. The RV of S2 measured with SPIFFIER in April and July 2022 is added and marked with arrow.}
  \label{fig:GC_spiffier} 
   \end{figure}

ERIS is an important next step in the Galactic center research. First of all, it will allow to continue simultaneous monitoring of a large number of stars orbiting around the SgrA$^*$ black hole (Fig.~\ref{fig:GC_spiffier}) and, therefore, continue to improve our knowledge of the distance and the mass of the black hole. In addition, higher Strehl ratios provided by the AOF as well as the increased throughput and instrumental line profiles of SPIFFIER, will lead to discovery of new fainter stars, which were not seen by SPIFFI. Finally, exotic events like gas clouds and flares will continue being observed by SPIFFIER to improve our understanding of the evolution of the Galactic Center and accretion processes onto its black hole. 

\subsection{High-redshift galaxies}

One of the main science drivers of ERIS is to map the distribution of star formation, physical conditions of the interstellar medium (ISM), and the motions within galaxies at redshift z~$\sim$~1--3. At this epoch, galaxies were forming stars most rapidly so it is a key epoch to study the early assembly of the bulk of their stellar mass. 

\begin{figure} %[ht]
   \begin{center}
   \includegraphics[width=0.55\textwidth]{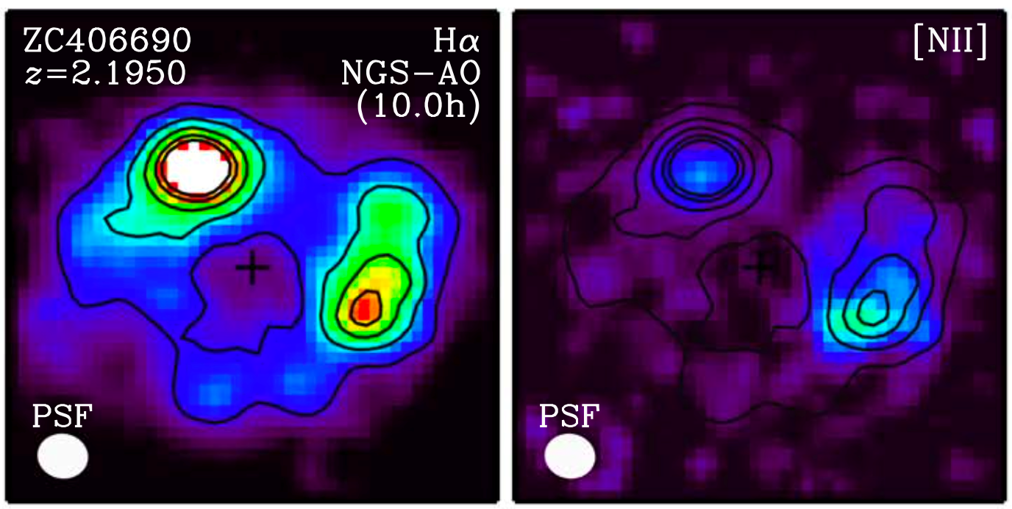}
   \includegraphics[width=0.55\textwidth]{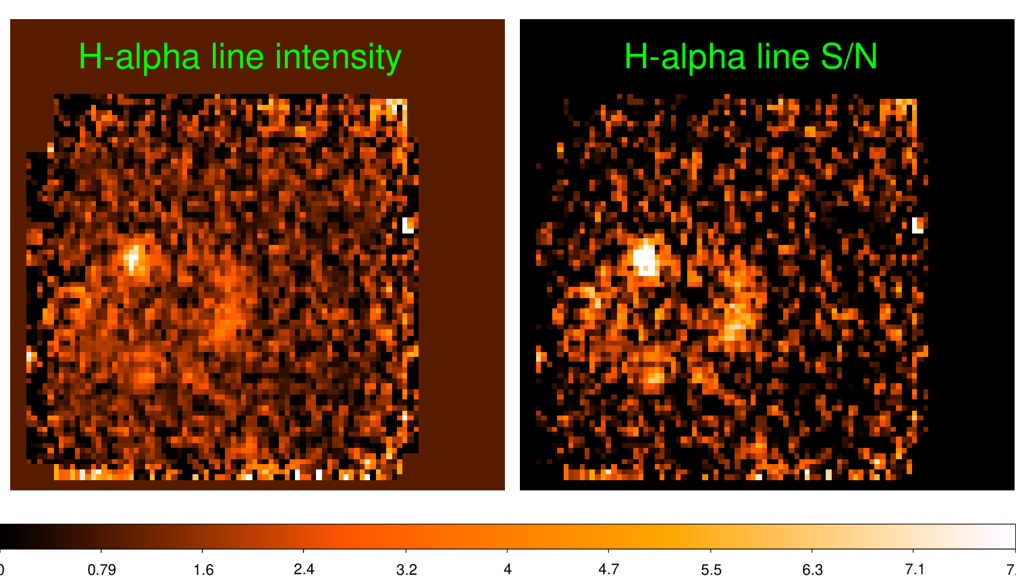}
   \end{center}
   \caption{\textit{Top panel:}  H-$\alpha$ and [NII] line intensity maps from SINFONI \cite{2018ApJS..238...21F}. The integration time was 10 hours on source. \textit{Bottom panel:} SPIFFIER reconstructed images in the H-$\alpha$ line; the spatial scale is 100mas/px. The integration time was 1~hour on source.}
  \label{fig:highz} 
\end{figure} 

The key capabilities of ERIS + AOF are the sensitivity, spectral resolution, and spatial resolution. The performance of AOF is far superior to that of the previous AO at UT4 with much higher Strehl ratios achievable. Along with improvements from the various SPIFFIER elements in terms of throughput, this leads to a sensitivity about 4-5 times higher than SINFONI + AO for compact structures on scales of $\sim$~0.1", or $\sim$~1~kpc at z~$\sim$~1--3. The better AO performance also allows to push the AO star to fainter magnitude substantially increasing the sky coverage. In addition, the improved spectral resolution of SPIFFIER (especially the R$\sim$10000 capability) is crucial to better constrain the kinematics, disentangle non-circular motions that are present within otherwise globally regular disk rotation kinematics, and measure disk velocity dispersion roughly twice lower than before. These are key enabling capabilities to study the inner workings of distant galaxies on the physically relevant Toomre scale (the characteristic fragmentation scale of globally unstable gas-rich disks as observed at high redshift), which is about 1 kpc at z~$\sim$~1--3. Science goals at the forefront of galaxy evolution research can now be addressed with ERIS systematically and for large samples, including gas transport within galaxies, the nature and fate of giant star-forming complexes, the origin of the elevated gas turbulence, the mass and energetics of feedback from star formation and active galactic nuclei (AGN), and the spatial distribution of gas-phase metallicity across galaxies.

The target observed during commissioning, zC406690, has been observed with SINFONI + AO with 10 hours on-source \cite{2018ApJS..238...21F}.  It is a typical z = 2.2 star-forming galaxy of $4 \times 10^{10} M_{\odot}$ and the star formation rate (SFR) of about 250 $M_{\odot}$/yr. A large fraction of this star formation takes place in kpc-scale "clumps" along a ring-like distribution (top panel of Fig.~\ref{fig:highz}).  These clumps drive powerful gas outflows, and this galaxy is one of the most striking example of star formation feedback. Ultimately, with longer integration, the higher spectral and spatial resolution will provide tighter constraints on the individual clump outflows and their impact on the surrouding ISM as well as on the clumps longevity. The purpose of the observations was to assess the performance of ERIS + AOF in realistic cases as will apply to faint distant galaxies, via a direct comparison with existing SINFONI+AO data. The bottom panel of Fig.~\ref{fig:highz} illustrates reconstructed SPIFFIER images in the H-$\alpha$ line at the 100 mas/px spatial scale: the morphology of the galaxy clumps and the diffuse emission along the ring are already recognized after about one hour of integration on source. This data set fully matches the expected angular resolution gain, and the sensitivity improvements for compact sources and extended emission of ERIS+AOF compared to SINFONI+AO.

\acknowledgments % equivalent to \section*{ACKNOWLEDGMENTS}       
 We would like to thank all the ESO staff in Garching and Paranal who have supported the installation and commissioning of ERIS, and to all consortium members who helped to design and build the instrument.

% References
\bibliography{report} % bibliography data in report.bib
\bibliographystyle{spiebib} % makes bibtex use spiebib.bst

\end{document}